\documentclass[]{article}
\usepackage{graphics,epsfig}
\title{The hidden measurement formalism: what can be explained and where quantum paradoxes remain\footnote{Published as:
Aerts, D., 1998, ``The hidden measurement formalism: what can be explained and where quantum paradoxes remain", {\it
International Journal of Theoretical Physics}, {\bf 37}, 291.}}
\author{Diederik Aerts}
\date{}
\font\smallroman=cmr10 at 8pt

\begin{document}
\maketitle
\centerline{FUND and CLEA,}
\centerline {Brussels Free University, Krijgskundestraat 33,}
\centerline {1160 Brussels, Belgium,}
\centerline {e-mail: diraerts@vub.ac.be}
\begin{abstract}
\noindent In the hidden measurement formalism that we develop in Brussels we explain the quantum
structure as due to the presence of two effects, (a) a  real change of state of the system under influence of
the measurement and, (b) a lack of knowledge about a deeper deterministic reality of the measurement process.
We show that the presence of these two effects leads to the major part of the quantum mechanical structure of
a theory describing a physical system where the measurements to test the properties of this physical system
contain the two mentioned effects.
We present a  quantum machine, where we can illustrate in a simple way how the quantum structure arises as a 
consequence of the two effects. We introduce a parameter $\epsilon$ that measures the amount of the
lack of knowledge on the measurement process, and by varying this parameter, we describe a continuous
evolution from a quantum structure (maximal lack of knowledge) to a classical structure (zero lack of
knowledge). We  show that for intermediate values of $\epsilon$ we find a new type of structure that is 
neither quantum nor classical. We analyze the quantum paradoxes in the light of these findings and show that
they can be divided into two groups: (1) The group (measurement problem and Schr{\"o}dingers cat paradox)
where the paradoxical aspects arise mainly from the application of standard quantum theory as a general
theory (e.g. also describing the measurement apparatus). This type of paradox
disappears in the hidden measurement formalism. (2) A second group collecting the paradoxes connected to the
effect of non-locality (the Einstein-Podolsky-Rosen paradox and the violation of Bell
inequalities). We show that these paradoxes are internally resolved because the effect of non-locality turns
out to be a fundamental property of the hidden measurement formalism itself.
\end{abstract}

\section{Introduction}
Quantum mechanics was originally introduced as a non commutative matrix calculus of observables
by Werner Heisenberg (Heisenberg 1925) and parallel as a wave mechanics by Erwin
Schr{\"o}dinger (Schr{\"o}dinger 1926). Both structurally very different theories, matrix mechanics and wave
mechanics could explain fruitfully the early observed quantum phenomena. Already in the same year the two
theories were shown to be realizations of the same, more abstract, ket-bra formalism by
Dirac (Dirac 1958). Only some years later, in 1934, John Von Neumann put forward a rigorous mathematical
framework for quantum theory in an infinite dimensional separable complex Hilbert space (Von Neumann 1955).
Matrix mechanics and wave mechanics appear as concrete realizations: the first one if the Hilbert space is
$l^2$, the collection of all square summable complex numbers, and the second one if the Hilbert space is
$L^2$, the collection of all square integrable complex functions. The formulation of quantum
mechanics in the abstract framework of a complex Hilbert space is now usually referred to as the `standard
quantum mechanics'.

The basic concepts  - the vectors of the Hilbert space representing the states of the system and the
self-adjoint operators representing the observables - in this standard quantum mechanics are abstract
mathematical concepts defined mathematically in and abstract mathematical space, and this is a problem
for the physicists working to understand quantum mechanics. Several
approaches have generalized the standard theory starting from more physically defined basic
concepts. John Von Neumann and Garett Birkhoff have initiated one of these approaches
(Birkhoff and Von Neumann 1936) were they analyze the difference between quantum and
classical theories by studying the `experimental propositions'. They could show that for a given
physical system classical theories have a Boolean lattice of experimental propositions while for quantum
theory the lattice of experimental propositions is not Boolean. Similar fundamental structural differences
between the two theories have been investigated by concentrating on different basic concepts. The collection
of observables of a classical theory was shown to be a commutative algebra while this is now the case for the
collection of quantum observables (Segal 1947, Emch 1984). Luigi Accardi and Itamar Pitowski obtained a
analogous result by concentrating on the probability models connected to the two theories: classical theories
have a Kolmogorovian probability model while the probability model of a quantum theory is non Kolmogorovian
(Accardi 1982, Pitowski 1989). The fundamental structural differences between the two types of theories,
quantum and classical, in different categories, was interpreted as indicating also a fundamental
difference on the level of the nature of the reality that both theories describe: the micro world should be
`very different' from the macro world. The author admits that he was himself very much convinced of this
state of affairs also because very concrete attempts to understand quantum mechanics in a classical way had
failed as well: e.g. the many `physical' hidden variable theories that had been tried out (Selleri 1990). In
this paper we want to concentrate on this problem: in which way the quantum world is different from the
classical world. We shall do this in the light of the approach that we have been elaborating in Brussels and
that we have called the `hidden measurement formalism'. We concentrate also on the different paradoxes in
quantum mechanics: the measurement problem, the Schr{\"o}dinger cat paradox, the classical limit, the
Einstein-Podolsky-Rosen paradox and the problem of non-locality. We investigate which ones of these quantum
problems are due to shortcomings of the standard formalism and which ones point out real physical differences
between the quantum and classical world.

\section{The two Major Quantum Aspects in Nature.}
As we mentioned already in the foregoing section, the structural difference between quantum
theories and classical theories (Boolean lattice versus non-Boolean lattice of propositions, commutative
algebra versus non commutative algebra of observables and Kolmogorovian versus non Kolmogorovian probability
structure) is one of the most convincing elements for the belief in a deep difference between the quantum
world and the classical world. During all the years that these structural differences have been investigated
(mostly mathematically) these has not been much understanding of the physical
meaning of these structural differences. In which way would these structural differences be linked to some
more intuitive but physically better understood differences between quantum theory and classical theory?

Within the hidden measurement approach we have been able to identify the physical aspects that are at the
origin of the structural differences between quantum and classical theories. This are two aspects that both
characterize the nature of the measurements that have to be carried out to test the properties of the system
under study. Let us formulate these two aspects carefully first. 

\medskip
\noindent
{\it We have a quantum-like theory describing a
system under investigation if the measurements needed to test the properties of the system are such that:}

\begin{description}

\item[(1)] {\it The measurements are not just observations but provoke a real change of the
state of the system}

\item[(2)] {\it There exists a lack of knowledge about the reality of what happens during the measurement
process}

\end{description}

\noindent It is the lack of knowledge {\bf (2)} that is theoretically structured in a non Kolmogorovian
probability model.In a certain sense it is possible to interpret the second aspect, the presence of the
lack of knowledge on the reality of the measurement process, as the presence of `hidden measurements' instead
of `hidden variables'. Indeed, if a measurement is performed with the presence of such a lack of knowledge,
then this is actually the classical mixture of a set of classical hidden measurements, were for such a
classical hidden measurement there would be no lack of knowledge. In an analogous way as in a hidden variable
theory, the quantum state is a classical mixture of classical states. This is the reason why we have called
the formalism that we are elaborating in Brussels and that consists in formalizing in a mathematical
theory the physical situations containing the two mentioned aspects, the `hidden measurement formalism'.

\section{Quantum Machine producing Quantum Structure}

\noindent After we had identified the two aspects {\bf (1)} and {\bf (2)} it was not difficult to invent a quantum
machine fabricated only with macroscopic materials and producing a quantum structure isomorphic to the
structure of a two dimensional complex Hilbert space, describing for example the spin of a quantum particle
with spin ${1 \over 2}$ (Aerts 1985, 1986, 1987). This quantum machine has been presented in different
occasions meanwhile (Aerts 1988a,b, 1991a, 1995) and therefore we shall only, for the sake of completeness,
introduce it shortly here.

The machine that we consider consists of a physical entity $S$ that is a point particle $P$ that can move on
the surface of a sphere, denoted $surf$, with center $O$ and radius $1$. The unit-vector $v$ where the
particle is located on $surf$ represents the state $p_v$ of the particle (see Fig. 1,a).  For each point $u \in
surf$, we introduce the following measurement $e_u$. We consider the diametrically opposite point $-u$, and
install a piece of elastic of length 2, such that it is fixed with one of its end-points in $u$ and the other
end-point in $-u$. 

\vskip 0.7 cm
\hskip 3 cm \includegraphics{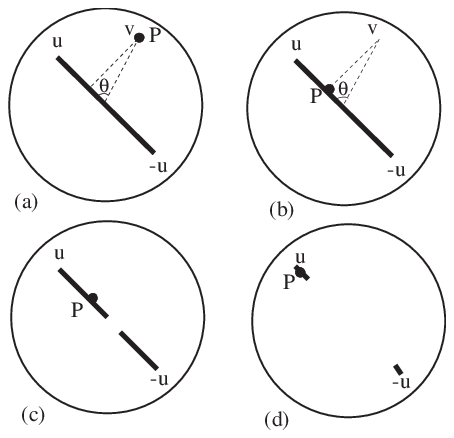}

\begin{quotation}
\noindent \baselineskip= 12 pt \smallroman Fig. 1 : A representation
of the quantum machine. In (a) the physical entity $\scriptstyle P$ is in state $\scriptstyle p_v$ in the
point $\scriptstyle v$, and the elastic corresponding to the measurement  $\scriptstyle e_{u}$ is installed
between the two diametrically opposed points
$\scriptstyle u$ and $\scriptstyle -u$. In (b) the particle $\scriptstyle P$ falls orthogonally onto the
elastic and stick to it. In (c) the elastic breaks and the particle $\scriptstyle P$ is pulled towards the
point $\scriptstyle u$, such that (d) it arrives at the point $\scriptstyle u$, and the measurement
$\scriptstyle e_u$ gets the outcome
$\scriptstyle o^u_1$.
\end{quotation}
Once the elastic is installed, the particle $P$ falls from its original place $v$
orthogonally onto the elastic, and sticks on it (Fig 1,b). Then the elastic breaks and the particle
$P$, attached to one of the two pieces of the elastic (Fig 1,c), moves to one of the two end-points $u$ or
$-u$ (Fig 1,d). Depending on whether the particle $P$ arrives in $u$ (as in Fig 1) or in $-u$, we give the
outcome $o^u_1$ or $o^u_2$ to $e_u$. We can easily calculate the probabilities corresponding to the two
possible outcomes. Therefore we remark that the particle $P$ arrives in $u$ when the  elastic breaks in a
point of the interval $L_1$ (which is the length of the  piece of the elastic between  $-u$ and the point
where the particle has arrived,  or  $1+cos\theta$) , and arrives in $-u$ when it breaks in a point of the
interval
$L_2$  ($L_2=L-L_1=2-L_1$). We make the hypothesis that the elastic breaks uniformly, which means that the
probability that the particle, being in state $p_v$, arrives in $u$, is given by the length of $L_1$ divided
by the length of the total elastic  (which is 2). The probability that the particle in state $p_v$ arrives in
$-u$ is the length of $L_2$ (which is $1-cos\theta$) divided by the length of the total elastic. If we denote
these probabilities respectively by $P(o^u_1, p_v)$ and $P( o^u_2, p_v)$ we have: 
$$P(o^u_1, p_v) = {{1+cos\theta}\over 2} = cos^2{\theta\over 2} \quad \quad \quad \quad P(o^u_2, p_v) =
 {{1-cos\theta}\over 2} = sin^2{\theta\over 2} \eqno(1)$$

\noindent These transition probabilities are the same as the ones related to the  outcomes of a Stern-Gerlach
spin measurement on a spin ${1 \over 2}$  quantum particle, of which the quantum-spin-state in direction $v =
(cos\phi sin\theta,$ $ sin\phi sin\theta, cos\theta)$, denoted by $\bar {\psi_v}$, and the measurement $e_u$
corresponding to the spin measurement in direction $u = (cos\beta sin\alpha,$ $sin\beta sin\alpha,
cos\alpha)$, is described respectively by the vector and the self adjoint operator of a two-dimensional
complex Hilbert space.
$$\psi_v = (e^{-i\phi/2}cos\theta/2, e^{i\phi/2}sin\theta/2)\ \quad \quad \quad  H_u = {1\over 2}
\pmatrix{cos\alpha & e^{-i\beta}sin\alpha \cr
e^{i\beta}sin\alpha &cos\alpha \cr} \eqno(2)$$
We can easily see now the two aspects in this quantum machine that we have identified in the hidden
measurement approach to give rise to the quantum structure. The state of the particle $P$ is effectively
changed by the measuring apparatus ($p_v$ changes to
$p_u$ or to $p_{-u}$ under the influence of the measuring process), which identifies the first aspect, and
there is a lack of knowledge on the interaction between the measuring apparatus and the particle, namely the
lack of knowledge of were exactly the elastic will break, which identifies the second aspect. We can also
easily understand now what is meant by the term `hidden measurements'. Each time the elastic breaks
in one specific point $\lambda$, we could identify the measurement process that is carried out afterwards as
a hidden measurement $e_u^\lambda$. The measurement $e_u$ is then a classical mixture of the collection of
all measurement $e_u^\lambda$: namely $e_u$ consists of choosing at random one of the $e_u^\lambda$ and
performing this chosen $e_u^\lambda$.

\section{The Quantum Classical Relation}

\noindent First of all we remark that we have shown in our group in Brussels that such a hidden measurement
model can be built for any arbitrary quantum entity (Aerts 1985, 1986, 1987, Coecke 1995a,b,c). However, the
hidden measurement formalism is more general than standard quantum theory. Indeed, it is very easy to
produce quantum-like structures that cannot be represented in a complex Hilbert space (Aerts 1986).

If the quantum structure can be explained by the  presence of a lack of knowledge on the measurement process,
we can go a step further, and wonder what types of structure arise when we consider the original models, with
a lack of knowledge on the measurement process, and introduce a variation of the magnitude of this lack of
knowledge.  We have studied the quantum machine under varying `lack of knowledge', parameterizing  this
variation by a number $\epsilon \in [0,1]$, such that $\epsilon = 1$ corresponds to the situation of maximal
lack of knowledge, giving rise to a quantum structure, and $\epsilon = 0$ corresponds to the situation of zero
lack of knowledge, generating a classical  structure. Other values of $\epsilon$ correspond
to intermediate situations and give rise to a  structure that is neither quantum nor classical (Aerts
Durt Van Bogaert 1992, 1993 and Aerts Durt 1994 a,b). We have called this model the
$\epsilon$-model and want to introduce it again in  this paper to explain in which way some of the quantum
paradoxes are solved in the hidden measurement formalism.

\bigskip
\par\noindent {\it a) The $\epsilon$-Model}
\par\noindent
We start from the quantum machine, but introduce now different types of elastic. An $\epsilon, d$-elastic
consists of three different parts: one  lower part where it is unbreakable, a middle part where it breaks
uniformly, and an upper  part where it is again unbreakable. By means of the two parameters $\epsilon \in
[0,1]$  and $d \in [-1+\epsilon, 1-\epsilon]$, we fix the sizes of the three parts in the following way.
Suppose that we have installed the $\epsilon, d$-elastic between the points $-u$ and $u$ of the sphere. Then
the elastic is unbreakable in the lower part from $- u$ to $(d-\epsilon) \cdot
u$, it breaks uniformly in the part from $(d-\epsilon) \cdot u$ to $(d+\epsilon) \cdot u$, and it is again
unbreakable in the upper part from $(d+\epsilon) \cdot u$ to $u$ (see Fig. 2).

\vskip 0.7 cm
\hskip 3 cm \includegraphics{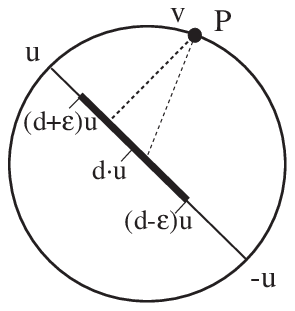}

\begin{quotation}
\noindent \baselineskip= 12 pt \smallroman Fig. 2 :  A representation of the measurement $\scriptstyle e^\epsilon_{u,
d}$. The elastic breaks uniformly between the  points $\scriptstyle (d-\epsilon) u$ and $\scriptstyle (d +\epsilon)
u$,  and is unbreakable in other points.
\end{quotation}
An $e_u$ measurement performed by means of an $\epsilon, d$-elastic shall be denoted by
$e^\epsilon_{u,d}$. We have the following cases:
\smallskip
\noindent {\it (1)} $v \cdot u \le d-\epsilon$. The particle sticks to the lower part of the $\epsilon,
d$-elastic, and any breaking of the elastic pulls  it down to the point $-u$. We have $P^\epsilon(o^u_1, p_v)
= 0$ and $P^\epsilon(o^u_2, p_v) = 1$.

\smallskip
\noindent {\it (2)} $d-\epsilon < v \cdot u < d+\epsilon$. The particle falls onto the breakable part of the
$\epsilon, d$-elastic. We can easily  calculate the transition probabilities and find:
$$P^\epsilon(o_1^u, p_v) = {1 \over 2\epsilon}( v \cdot u -d +\epsilon) \quad \quad \quad P^\epsilon(o_2^u,
p_v) = {1 \over 2\epsilon}(d+\epsilon - v \cdot u) \eqno(3)$$
\smallskip
\noindent {\it (3)} $d+\epsilon \le v \cdot u$. The particle falls onto the upper part of the
$\epsilon,d$-elastic, and any  breaking of the elastic pulls it upwards, such that it arrives in $u$. We have 
$P^\epsilon(o_1^u, p_v) = 1$ and $P^\epsilon(o_2^u,  p_v) = 0$.

\smallskip
\noindent
 We are now in a very interesting situation from the point of view of the structural studies of quantum
mechanics. Since the $\epsilon$-model describes a continuous transition from quantum to classical, its
mathematical structure should be able to learn us {\it what are the structural shortcomings of the standard
Hilbert space quantum mechanics}. Therefore we have studied the $\epsilon$-model in the existing mathematical
approaches that are more general than the standard quantum mechanics: the lattice approach, the probabilistic
approach and the *-algebra approach.

\bigskip
\par\noindent {\it b) The Lattice Approach}
\par\noindent In this lattice approach exists a well known axiomatic scheme that reduces the approach to
standard quantum mechanics if certain axioms are fulfilled. For intermediate values of
$\epsilon$, that is $0 <  \epsilon < 1$, we find that 2 of the 5 axioms needed in the lattice approach
to reconstruct standard quantum mechanics are violated. The axioms that are violated are the weak-modularity
and the covering law, and it are precisely those axioms that are needed to recover the vector space structure
of the state space in quantum mechanics (Aerts Durt Van Bogaert 1993 and Aerts Durt 1994 a,b). 

\bigskip
\par\noindent {\it  c) The Probabilistic Approach}
\par\noindent If we take the case of vanishing fluctuations ($\epsilon = 0$), do we obtain the Kolmogorovian
theory of probability? This would be most interesting, since then we would have constructed a macroscopic
model with an understandable structure (i.e., we can see how the probabilities arise) and  a quantum and a
classical behavior. We (Aerts 1995) proposed to test the polytopes for a
family of conditional probabilities. The calculations can be found in (Aerts S. 1995) and
the result was what we hoped for: a macroscopic model with a quantum and a Kolmogorovian limit. For
intermediate values of the fluctuations ($0 <  \epsilon < 1$) the resulting probability model is
neither quantum nor Kolmogorovian: we have identified here a new type of probability model, that is
quantum-like, but not really isomorphic to the probability model found in a complex Hilbert space.

\bigskip
\par\noindent{\it d) The *-Algebra Approach}
\par\noindent The *-algebra provides a natural mathematical language for quantum
mechanical operators.  We applied the concepts of this approach to the epsilon model  
to find that an operator corresponding to an $\epsilon$ measurement is linear if and only if $\epsilon = 1$
(Aerts D'Hooghe 1996). This means that for the classical and intermediate situations the observables cannot
be described by linear operators.

\bigskip
\noindent {\bf Conclusion:} {\it Quantum theory and classical theories appear as special cases ($\epsilon = 1,
\epsilon = 0$) and the general intermediate case, although quantum-like, cannot be described in standard
Hilbert space quantum mechanics.}

\bigskip \noindent {\it e) The Measurement Problem and the Schr{\"o}dingers Cat Paradox}
\par \noindent The result stated in the conclusion means that all the paradoxes of standard quantum theory
that are due to the fact that quantum theory is used as a universal theory, also being applied to
macroscopic system, for example the measuring apparatus, are not present in our hidden measurement
formalism. We explicitly have in mind the `measurement problem' and the `Schr{\"o}dinger cat paradox'.
Indeed, the measurement apparatus should be described by a classical model in our approach, and the physical
system eventually by a quantum model. The problem of the presence of quantum correlations between physical
system and measuring apparatus, as it presents itself in the standard theory, takes a completely different
aspect. We are working now at the elaboration of a concrete description of the measurement process within the
hidden measurement formalism (Aerts Durt 1994b). Our result also shows that it is possible in the hidden
measurement formalism to formulate a `classical limit', namely as a continuous transition from quantum to
classical (Aerts Durt van Bogaert 1993).

\section{Non-Locality as a Genuine Property of Nature}

The measurement problem and paradoxes equivalent to Schr{\"o}dingers cat paradox disappear in
the hidden measurement formalism, because standard quantum mechanics appears only as a special case, the
situation of maximum fluctuations. All quantum mysteries connected to the effect of `non-locality' remain
(Einstein Podolsky Rosen paradox and the violation of the Bell inequalities). It is even so that non-locality
unfolds itself as a fundamental aspect of the hidden measurement approach. This is due to the fact that if we
explain the quantum structures as it is done in the hidden measurement formalism a quantum measurement
has two concrete physical effects: (1) it changes the state of the system and (2) it produces probability
due to a lack of knowledge about the nature of this change. With the quantum machine we have
given a macroscopic model for a spin measurement of a spin ${1
\over 2}$ particle. If we apply the hidden measurement formalism to the situation of a quantum system
described by a wave function $\psi(x)$, and to a position or a momentum measurement performed on this
system, we also have the two mentioned effects. For example in the case of a position measurement: the
detection apparatus changes the state of the quantum system, in the sense that it `localizes' the quantum
system in a specific place of space, and the probabilities that are connected with this measurement are due
to the fact that we lack knowledge about the specific way in which this localization takes place. This means
that `before the detection has taken place' the quantum system was in general not localized: it was not
present in a specific region of space. With other words, quantum systems are fundamentally non-local systems: the wave function
$\psi(x)$ describing such a quantum state is not interpreted as a wave that is present in space, but
as indicating these regions of space were the particle can be localized, where 
$\int_R|\psi(x)|^2dx$ is the probability that this localization will happen in region
$R$. A similar interpretation must be given to the momentum measurement of a quantum entity: the quantum
entity has no momentum before the measurement, but the measurement creates partly this momentum.
 In (Aerts Durt Van Bogaert 1993) we have calculated the $\epsilon$-situation for a quantum system described
by a wave function $\psi(x)$, element of the Hilbert space of all square integrable complex
functions, and we have found the following very simple procedure.

\vskip 0.7 cm
\hskip 3 cm \includegraphics{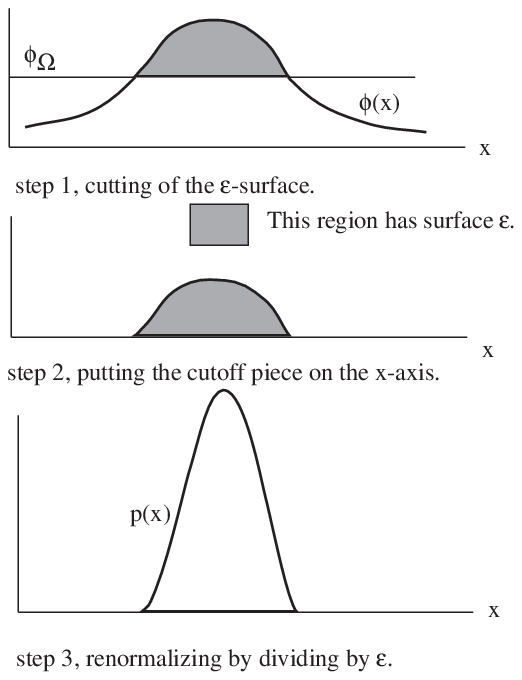}

\begin{quotation}
\noindent \baselineskip= 12 pt \smallroman Fig. 3 : A graphical representation of the $\scriptstyle \epsilon$ situation
for a quantum system described by a wave function $\scriptstyle \psi(x)$. Suppose that $\scriptstyle \epsilon \in
[0,1]$ is given and $\scriptstyle \phi(x) = |\psi(x)|^2$ is the probability distribution corresponding to $\scriptstyle
\psi(x)$. We cut, by means of a
constant function $\scriptstyle \phi_\Omega$, a piece of the function $\scriptstyle \phi(x)$, such that the surface
contained in the cutoff piece equals $\scriptstyle \epsilon$ (step 1 of fig 3). We move this piece of function to the
$\scriptstyle x$-axis (step 2 of fig 3), and then renormalize by dividing by $\scriptstyle \epsilon$ (step 3 of
fig 3).
\end{quotation}
Suppose that $\epsilon$ is given, and the
state of the quantum system is described by the wave function $\psi(x)$ and 
$\phi(x)$ is the corresponding probability distribution (hence $\phi(x) = |\psi(x)|^2$). We cut, by means of a
constant function $\phi_\Omega$, a piece of the function $\phi(x)$, such that the surface contained in the
cutoff piece equals $\epsilon$ (see step 1 of fig 3). We move this piece of function to the $x$-axis (see
step 2 of fig 3). And then we renormalize by dividing by $\epsilon$ (see step 3 of fig 3). If we
proceed in this way for smaller values of $\epsilon$, we shall finally arrive at a delta-function for
the classical limit $\epsilon \to 0$, and the delta-function is located in the original maximum of
the quantum probability distribution. We want to point out that the state $\psi(x)$ of the physical system
is not changed by this $\epsilon$-procedure, it remains always the same state, representing the same
physical reality. It is the regime of lack of knowledge going together with the detection measurement that
changes with varying $\epsilon$. For $\epsilon =1$ this regime is one of maximum lack of knowledge on the
process of localization, and this lack of knowledge is characterized by the spread of the probability
distribution $\phi(x)$. For an intermediate value of $\epsilon$, between 1 and 0, the spread of the
probability distribution has decreased (see fig3) and for zero fluctuations the spread is 0. 

Let us also try
to see what becomes of the non-local behavior of quantum entities taking into account the classical limit
procedure that we propose. Suppose that we consider a double slit experiment, then the state $p$ of a quantum
entity having passed the slits can be represented by a probability function $p(x)$ of the form represented in
Fig 4. We can see that the non-locality presented by this probability function gradually disappears when
$\epsilon$ becomes smaller, and in the case where $p(x)$ has only one maximum finally disappears
completely.

\vskip 0.7 cm
\hskip 3 cm \includegraphics{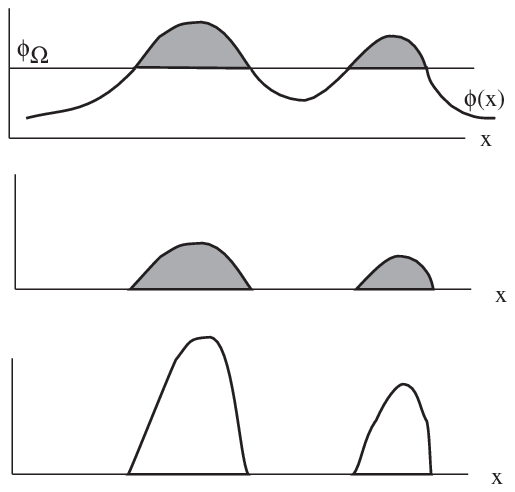}

\begin{quotation}
\noindent \baselineskip= 12 pt \smallroman Fig. 4 : A representation of the probability distribution corresponding to
the state of a quantum system that has passed a double slit. We show the three different steps of the $\scriptstyle
\epsilon$ procedure (Fig 3) in this case.
\end{quotation}
When there are no fluctuations on the measuring apparatus used to detect the
particle, it shall be detected with certainty in one of the slits, and always in the same one. If
$p(x)$ has two maxima (one behind slit 1, and the other behind slit 2) that are equal, the
non-locality does not disappear. Indeed, in this case the limit-function is the sum of two
delta-functions (one behind slit 1 and one behind slit 2). So in this case the non-locality remains
present even in the classical limit. If our procedure for the classical limit is a correct one, also
macroscopic classical entities can be in non-local states. How does it come that we don't
find any sign of this non-locality in the classical macroscopic world? This is due to the
fact that the set of states, representing a situation where the probability function has more
than one maximum, has measure zero, compared to the set of all possible states, and moreover
these states are `unstable'. The slightest perturbation will destroy the symmetry of the
different maxima, and hence shall give rise to one point of localization in the classical
limit. Also classical macroscopic reality is non-local, but the local model that we use to describe it
gives the same statistical results, and hence cannot be distinguished from the non-local model.

\section{The Violation of Bell Inequalities: Quantum, Classical and Intermediate}

\noindent It is interesting to consider the violation of Bell's inequalities within the hidden-measurement
formalism. The quantum machine, as presented in section 3, delivers us a macroscopic model for the spin of a
spin ${1 \over 2}$ quantum entity and starting with this model it is possible to construct a macroscopic
situation, using two of these models coupled by a rigid rod, that represents faithfully the situations of two
entangled quantum particles of spin ${1 \over 2}$ (Aerts 1991b). The `non-local' element is introduced
explicitly by means of a rod that connects the two sphere-models.  We also have studied this EPR situation of
entangled quantum systems by introducing the $\epsilon$-variation of the amount of lack of knowledge on the
measurement processes and could show that one violates the Bell-inequalities even more for classical but
non-locally connected systems, that is,
$\epsilon=0$. This illustrates  that the violation of the Bell-inequalities is due to the non-locality rather
then to the indeterministic character of quantum theory. And that the quantum indeterminism (for values of
$\epsilon$ greater than 0) tempers the violation of the Bell inequalities (Aerts D., Aerts S., Coecke and
Valckenborgh 1996, Aerts, Coecke, Durt and Valckenborgh, 1997a,b). This idea has been used to construct a
general representation of entangled states (hidden correlations) within the hidden measurement formalism (Coecke 1996, 1998
and Coecke, D'Hooghe and Valckenborgh 1997).

\section{References}

\begin{description}

\item Accardi, L., 1982, {\it Nuovo Cimento}, {\bf 34}, 161.

\item Accardi, L., 1984, ``The probabilistic roots of the quantum mechanical paradoxes", in {\it The
Wave-Particle Dualism}, eds. Diner, S. et al., Reidel Publishing Company, Dordrecht.

\item Aerts, D., 1982, ``Example of a macroscopical situation that violates Bell inequalities", {\it Lett. Nuovo
Cimento}, {\bf 34}, 107. 

\item Aerts, D., 1984, ``How do we have to change quantum mechanics in order to describe separated
systems", in {\it The Wave-Particle Dualism}, eds. Diner, S. et al., Reidel Publishing Company, Dordrecht. 

\item Aerts, D., 1985, ``A possible explanation for the probabilities of quantum mechanics and a
macroscopic situation that violates Bell inequalities", in {\it Recent Developments in Quantum Logic},
eds. Mittelstaedt, P. et al., in {\it Grundlagen der Exacten Naturwissenschaften}, vol.6, Wissenschaftverlag,
Bibliographisches Institut, Mannheim, 235.

\item Aerts, D., 1986, ``A possible explanation for the probabilities of quantum mechanics", {\it J. Math. Phys.} {\bf
27}, 202.

\item Aerts, D., 1987, ``The origin of the non-classical character of the
quantum probability model", in {\it Information, Complexity, and Control in Quantum Physics}, eds. Blanquiere,
A. et al., Springer-Verlag.

\item Aerts, D., 1988a, ``The physical origin of the EPR paradox and how to violate Bell
inequalities by macroscopic systems", in {\it Symposium on the Foundations of Modern Physics},
eds. Lahti, P. et al., World Scientific, Singapore.

\item Aerts, D., 1988b, ``The description of separated systems and a possible explanation for the
probabilities of quantum mechanics", in {\it Microphysical Reality and Quantum Formalism}, eds. van der
Merwe et al., Kluwer Academic Publishers.

\par \item Aerts, D., 1991a, ``A macroscopic classical laboratory situation with
only macroscopic classical entities giving rise to a quantum mechanical probability model",
in {\it Quantum Probability and Related Topics, Vol. VI,} ed. Accardi, L., World Scientific, Singapore.

\item Aerts, D., 1991b, ``A mechanistic classical laboratory situation
violating the Bell inequalities with 2$\sqrt{2}$, exactly `in the same way' as
its violations by the EPR experiments", {\it Helv. Phys. Acta}, {\bf 64}, 1. 

\item Aerts, D., 1994, ``Quantum structures, separated physical entities and
probability", {\it Found. Phys.}, {\bf 24}, 1227.

\item Aerts, D., 1995, ``Quantum structures: an attempt to explain their
appearance in nature", {\it Int. J. Theor. Phys.}, {\bf 34}, 1165.

\item Aerts, D., Aerts, S., Coecke, B. and Valckenborgh, F., 1996, ``The meaning of the violation
of Bell inequalities: non-local correlation or quantum behavior?", Preprint, CLEA, Brussels Free University,
Krijgskundestraat 33, Brussels.

\item Aerts, D., Coecke, B., Durt, T. and Valckenborgh. F., 1997a,
``Quantum, classical and intermediate I: A model on the poincar{\'e} sphere", {\it Tatra
Mountains Mathematical Publications}, {\bf 10}, 225.

\item Aerts, D., Coecke, B., Durt, T. and Valckenborgh. F., 1997b,
``Quantum, classical and intermediate II: The vanishing vector space structure",
{\it Tatra Mountains Mathematical Publications}, {\bf 10}, 241.

\item Aerts, D. and D'Hooghe, B., 1996, ``Operator structure of a
non-quantum and a non-classical system", {\it Int. J. Theor.
Phys.}, {\bf 35}, 2241.

\item Aerts, D., and Durt, T., 1994a, ``Quantum, classical and intermediate, an
illustrative example", {\it Found. Phys.}, {\bf 24}, 1353.

\item Aerts, D. and Durt, T., 1994b, ``Quantum, classical and intermediate: A
measurement model", in {\it 70 Years of Matter-Wave}, eds. Laurikainen, K.V.,
Montonen, C. and Sunnarborg, K., Editions Frontieres, Gives Sur Yvettes, France.

\item Aerts, D., Durt T. and Van Bogaert, B., 1993, ``A physical example of quantum fuzzy
sets, and the classical limit", {\it Tatra Mountains Mathematical Publications}, {\bf 5}, 15. 

\item Aerts, D.,  Durt, T. and Van Bogaert, B., 1993, ``Quantum probability, the classical limit and
non-locality",  in {\it On the
Foundations of Modern Physics 1993}, ed. Hyvonen, T., World
Scientific, Singapore, 35.

\item Aerts, S., 1996, ``Conditional probabilities with a quantal and a Kolmogorovian limit", {\it Int. J. Theor. Phys.},
{\bf 35}, 2245.

\item Bell, J., 1964, {\it Physics}, {\bf1}, 195.

\item Bohm, D., 1951, {\it Quantum Theory}, Prentice-Hall.

\item Birkhoff, G. and Von Neumann, J., 1936, ``The logic of quantum mechanics", {\it J., Ann. Math.}, 
{\bf 37}, 823.

\item Coecke, B., 1995a, ``Hidden measurement representation for quantum entities described by finite dimensional complex
Hilbert spaces", {\it Found. Phys.}, {\bf 25}, 1185. 
                    
\item Coecke, B., 1995b, ``Generalisation of the proof on the existence of hidden measurements to experiments with an
infinite set of outcomes", {\it Found. Phys. Lett.}, {\bf 8}, 437.

\item Coecke, B., 1995c, ``Hidden measurement model for pure and mixed states of quantum physics in Euclidean space", {\it
Int. J. Theor. Phys.}, {\bf 34}, 1313.

\item Coecke, B., 1996, ``Superposition states through correlation's of the second kind", {\it Int. J. Theor. Phys.},
{\bf 35}, 2371.

\item Coecke, B., 1998, ``A Representation for compound systems as individual entities: Hard acts of creation and hidden
correlations", {\it Found. Phys.}, {\bf 28}, 1109.

\item Coecke, B., D'Hooghe, B. and Valckenborgh, F., 1997, ``Classical physical entities with a quantum
description", in {\it Fundamental Problems in Quantum Physics II}, eds. Ferrero, M.
and van der Merwe, A., Kluwer Academic, Dordrecht, 103. 

\item Dirac, P.A.M., 1958, {\it The Principles of Quantum Mechanics}, Clarendon Press, Oxford.

\item Einstein, A., Podolsky, B. and Rosen, N., 1935, {\it Phys. Rev.}, {\bf 47}, 777.

\item Emch, G. G., 1984, {\it Mathematical and Conceptual Foundations of 20th Century
Physics,} North-Holland, Amsterdam.

\item Heisenberg, W., 1925, {\it Zeitschr. Phys.} {\bf 33}, 879.

\item Pitovski, I., 1989, {\it Quantum Probability - 
Quantum Logic}, Springer Verlag.

\item Schr{\"o}dinger, E., 1926, {\it Ann. der Phys.} {\bf 79}, 1926.

\item Segal, I. E., 1947, {\it Ann. Math.}, {\bf 48}, 930.

\item Selleri, F., 1990, {\it Quantum Paradoxes and Physical Reality}, Kluwer Academic Publishers,
Dordrecht.

\item Von Neumann, J., 1955, {\it Mathematical Foundations of Quantum Mechanics}, Princeton University
Press, Princeton.

\end{description}

\end{document}